\documentclass[twocolumn]{aastex62}

\def\lapp{\ifmmode\stackrel{<}{_{\sim}}\else$\stackrel{<}{_{\sim}}$\fi}
\def\gapp{\ifmmode\stackrel{>}{_{\sim}}\else$\stackrel{>}{_{\sim}}$\fi}
\usepackage{multirow}
\usepackage{color}
\usepackage{amsmath}
\usepackage{soul}
\usepackage{hyperref}

\shorttitle{}
\shortauthors{}
\begin{document}

\title{Characterizing X-ray properties of the gamma-ray pulsar PSR~J1418$-$6058 in the Rabbit pulsar wind nebula}

\correspondingauthor{Hongjun An}
\email{hjan@cbnu.ac.kr}

\author{Minjun Kim}
\author{Hongjun An}
\affiliation{Department of Astronomy and Space Science,\\
Chungbuk National University, Cheongju, 28644, Republic of Korea}

\begin{abstract}
        We report on X-ray studies of the gamma-ray pulsar PSR~J1418$-$6058 in
the Rabbit pulsar wind nebula (PWN) carried out using archival {\it Chandra}
and {\it XMM-Newton} observations.
A refined timing analysis performed with the 120-ks {\it XMM-Newton}
data finds significant ($p\approx10^{-7}$) pulsation at $P\approx110$\,ms which
is consistent with that measured with the {\it Fermi} large area
telescope (LAT).
In the {\it Chandra} image, we find extended emission around the pulsar
similar to those seen around other pulsars in young PWNe,
which further argues for association between PSR~J1418$-$6058 and the Rabbit PWN.
The X-ray spectrum of the pulsar is hard and similar to those of
soft-gamma pulsars. Hence PSR~J1418$-$6058 may add to the list of
soft-gamma pulsars.

\end{abstract}


\section{Introduction}
\label{sec:intro}
	Pulsars, rapidly spinning neutron stars, are one of main sources
of gamma-ray emission in the Galaxy, and so a large fraction of the {\it Fermi}
large area telescope \citep[LAT;][]{fermimission} sources is pulsars.
They are further categorized into several classes based on
their temporal and spectral properties, from rapidly-spinning
recycled pulsars to slowly-spinning hot magnetars \citep[e.g.,][]{harding13}.
Emission properties of pulsars are diverse; magnetars emit almost
exclusively in the X-ray band \citep[e.g.,][]{kb17}, and recycled pulsars mainly do in the radio and the
gamma-ray bands. Some relatively young pulsars emit nonthermal radiation
in the X-ray band which becomes softer at gamma-ray energies \citep[][]{kh15}.
Pulsars' timing properties are also various; the spin periods range from
milliseconds to tens of seconds, and some pulsars exhibit frequent timing anomaly
(e.g., glitches) while others do not. These observational diversities can
be explained with the fundamental neutron star physics to some degrees,
but more studies are needed to understand them better.
In addition, characterizing pulsar properties is important to understand
pulsar wind nebulae (PWNe) with which we can study the relativistic
shock physics \citep[][]{skl15} and magnetohydrodynamic flow
of high-energy particles \citep[][]{kc84a},

	Observational studies of pulsars are mainly done
in the radio, X-ray and gamma-ray bands \citep[e.g.,][]{ctlv19} because pulsars are bright in these bands.
Energetic gamma-ray pulsars are often discovered by {\it Fermi}-LAT blind searches
which are later confirmed by observations in other wavebands.
While continuous monitoring of the sky by {\it Fermi}
LAT allowed precise measurements of pulsars' spin properties \citep[][]{fermi2PC},
it is difficult to characterize their short-timescale properties
due to low count statistics in the gamma-ray band.
These can be measured better in the other wavebands, but
identifying a low-energy counterpart of a gamma-ray pulsar based on the position
is difficult due to insufficient angular resolution of gamma-ray telescopes; there
can be several X-ray point sources in the LAT-determined position error circle.
However, it was shown that {\it Fermi} timing analyses could determine
pulsars' position to high precision (to the arcsecond level),
which has been verified by previous studies \citep[e.g.,][]{krjs+15}.

\begin{figure*}
\centering
\hspace{-5.0 mm}
\begin{tabular}{cc}
\includegraphics[width=3.22 in]{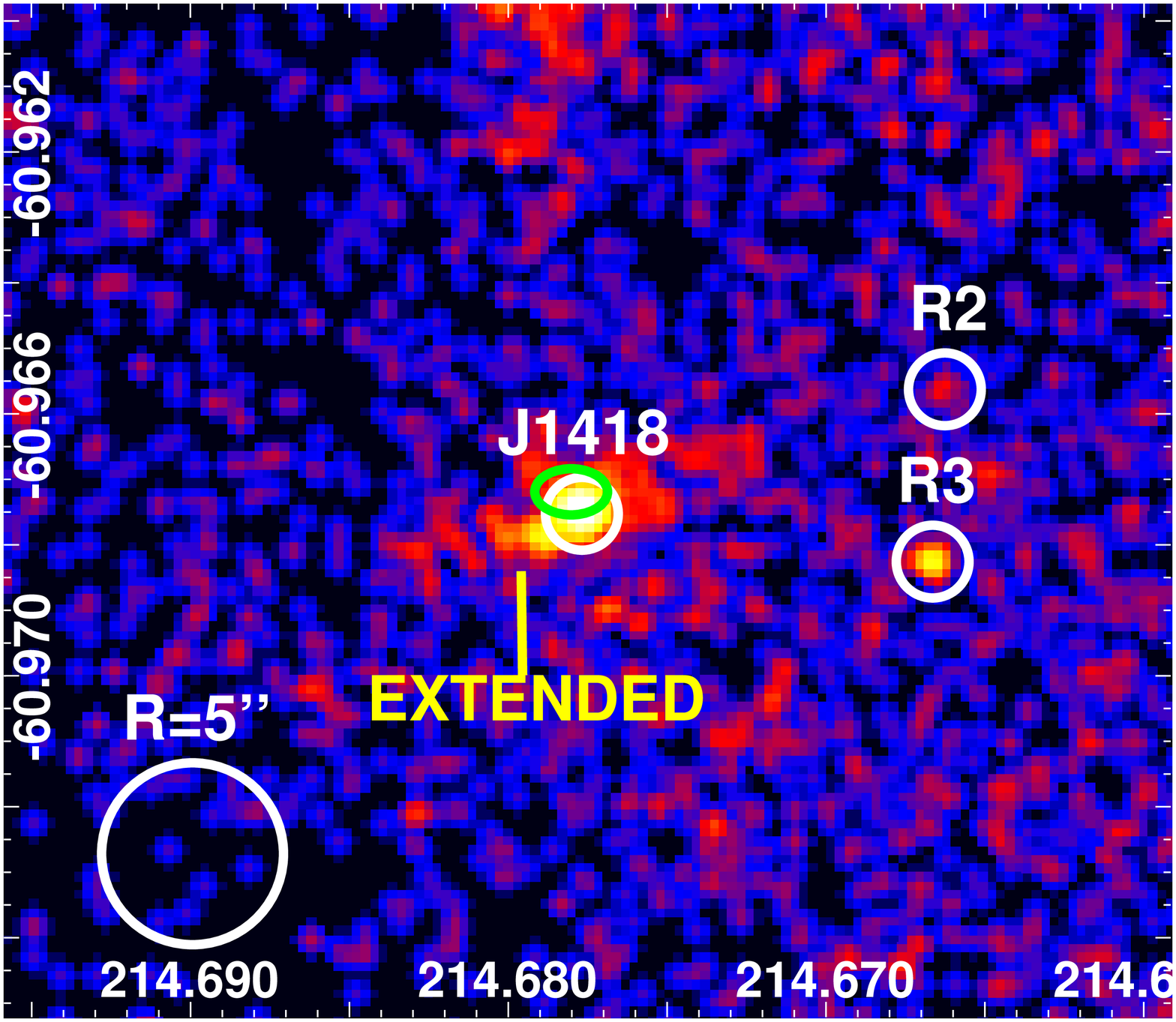} &
\includegraphics[width=3.22 in]{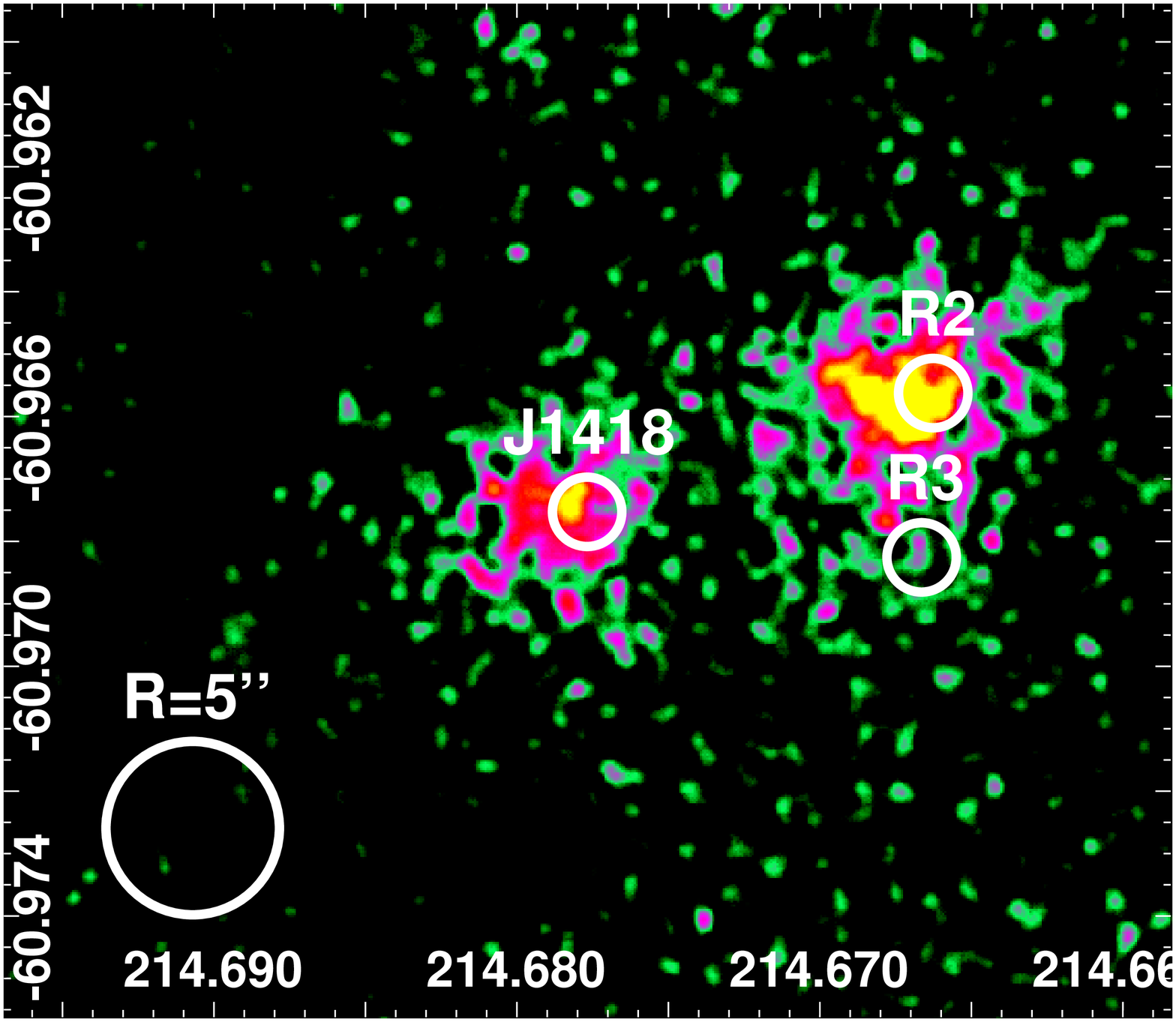} \\
\end{tabular}
\figcaption{1--10\,keV {\it Chandra} (left) and {\it XMM-Newton} Mos-combined (right) images 
of $\sim$$1'\times 1'$ regions around J1418.
Some sources are denoted with circles ($R=2''$), and an $R=5''$ circle
is shown in the bottom left corner for reference.
The green ellipse denotes the LAT-timing position of PSR~J1418$-$6058 \citep[][]{krjs+15}, and
extended emission around J1418 is clearly visible in the left panel.
Figures are smoothed and scales are adjusted for better legibility.
\label{fig:fig1}
}
\vspace{0mm}
\end{figure*}
	
	The pulsar PSR~J1418$-$6058 was first identified
as a bright gamma-ray source by {\it EGRET} \citep[][]{lm97}.
Follow-up radio and X-ray studies
found complex structure (Kookaburra region) around the source, most importantly
an extended radio and X-ray PWN \citep[``Rabbit'';][]{rrjg99}.
Interestingly extended TeV emission was detected near the PWN \citep[][]{hessrabbit2006} and
suggested to be associated with it.
However, it was unclear whether PSR~J1418$-$6058 is associated with the PWN
because there are multiple X-ray sources
in the region \citep[Fig.~\ref{fig:fig1}; see also][]{nrr05},
and X-ray pulsation was not detected significantly in any of them.
Subsequent timing studies with {\it Fermi} LAT
discovered a pulsar and allowed high-precision astrometry for the pulsar position
\citep[green circle in Fig.~\ref{fig:fig1} left;][]{krjs+15}
which is consistent with the position of an X-ray source. Hence the source (J1418 hereafter)
is likely to be the X-ray counterpart of PSR~J1418$-$6058.
Further X-ray studies will then help to measure
timing properties on short timescales and understand the nature of the pulsar.

	Moreover, if J1418 is associated with the Rabbit PWN,
characterizing J1418's properties can help to understand the intriguing PWN
having the TeV emission offset from the radio and X-ray one;
by modeling the X-ray and gamma-ray
light curves of the pulsar \citep[][]{hma98,rw10} 
and the torus morphology \citep[][]{nr04}, the pulsar emission geometry can be
inferred, which can then tell us about its energy injection to the PWN.

	In this paper, we investigate temporal, spectral, and spatial properties
of X-ray sources in the Rabbit PWN to identify unambiguously
the central pulsar (likely J1418) and to measure its properties using archival
{\it Chandra} and {\it XMM-Newton} data.

\section{Observational Data and Analysis}
\label{sec:sec2}

\subsection{Data reduction}
\label{sec:sec2_1}
       We use archival X-ray data obtained with the {\it Chandra} and
the {\it XMM-Newton} satellites. The data were taken on 2007 June 14
for 70\,ks (Obs. ID 7640) and on 2009 Feb. 21 for 120\,ks (Obs. ID 0555700101, MJD~54883)
with {\it Chandra} and {\it XMM-Newton}, respectively.
These data were reduced with the pipeline tools along
with the most recent calibration database for each observatory:
CIAO~4.11 and SAS~20180620. We processed the {\it Chandra} data
using the {\tt chandra\_repro} task and
the {\it XMM-Newton} data using the {\tt emproc} and {\tt epproc}
tasks with standard filters.
We further cleaned the {\it XMM-Newton} data to remove
particle-flare events using the standard flare-removal
procedure.\footnote{https://www.cosmos.esa.int/web/xmm-newton/sas-thread-epic-filterbackground}

\subsection{X-ray timing analysis}
\label{sec:sec2_2}

\begin{figure*}
\centering
\begin{tabular}{ccc}
\hspace{-3 mm}
\includegraphics[width=2.25 in]{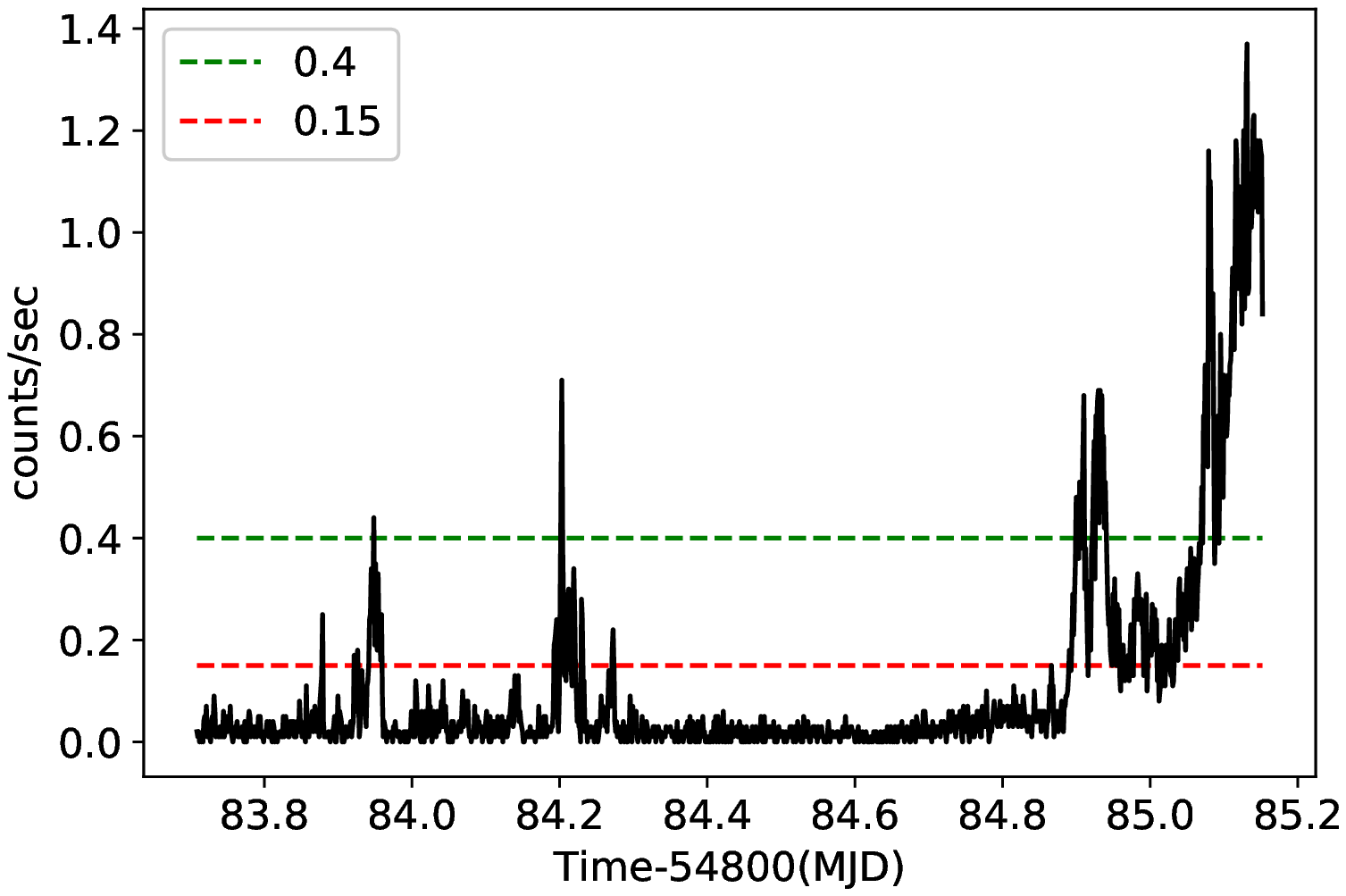} &
\includegraphics[width=2.25 in]{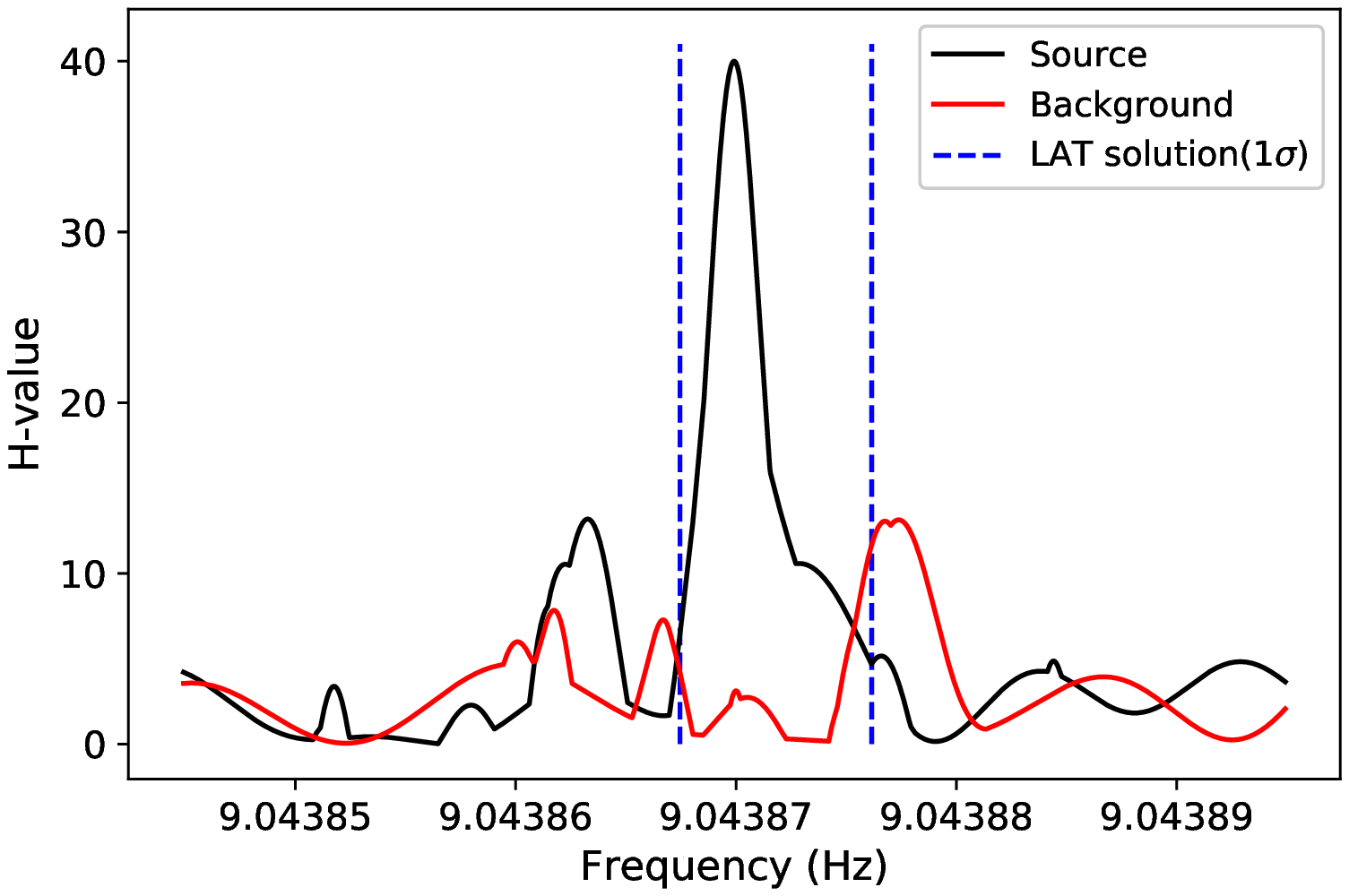} &
\includegraphics[width=2.25 in]{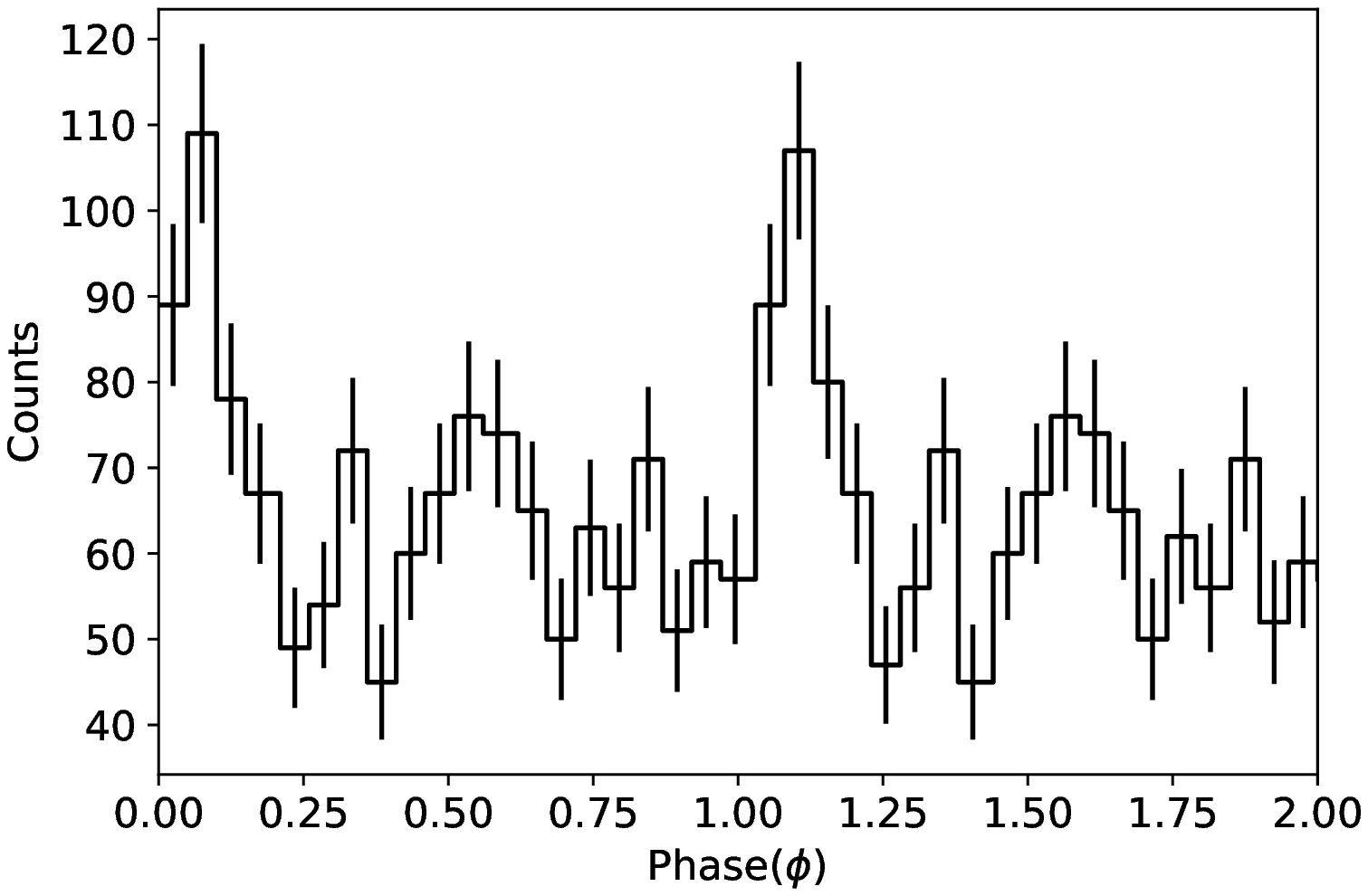} \\
\end{tabular}
\figcaption{{\it Left}: A particle-flare light curve measured with the {\it XMM-Newton} PN data.
Horizontal lines denote the flare filtering criteria:
a standard cut 0.4$\rm \ s^{-1}$ (green) and a tight cut 0.15$\rm \ s^{-1}$ (red).
{\it Middle}: $H$-test results for the source (black) events in the 1--10\,keV band, with
the flare cut of 0.15$\rm \ s^{-1}$.
Results for the background events (an $R=30''$ circle) are also shown in red and the 1$\sigma$
 range of the LAT solution at the X-ray epoch is shown in blue for reference.
{\it Right}: A 1--10\,keV pulse profile (flare cut 0.15$\rm \ s^{-1}$) constructed
by folding the source events on the best period (Section~\ref{sec:sec2_2}).}
\label{fig:fig2}
\vspace{0mm}
\end{figure*}

	We first search the X-ray data for pulsations (e.g., $P$=110\,ms)
of sources in the Rabbit PWN.
This can be done only with the {\it XMM-Newton}/PN small-window mode observation
with sufficient timing resolution ($\Delta T=5.7$\,ms; for a detection of the 110\,ms pulsation).
In the {\it XMM-Newton}/PN data, we find two bright sources $\sim$20$''$ away
from each other (Fig.~\ref{fig:fig1} right); one of them (the eastern one; J1418)
is positionally coincident with the gamma-ray pulsar PSR~J1418$-$6058
(green circle in the left panel).
Note that the point source denoted as R3 in the left panel ({\it Chandra} data)
is not present in the right panel ({\it XMM-Newton} data).

	We extract 1--10\,keV events within 10$''$ circles centered at the positions of
the two X-ray sources (J1418 and R2), and barycenter-correct the event arrival times using
(R.A., decl.)=($214.677695^\circ$, $-60.967483^\circ$)
and ($214.667428^\circ$, $-60.965500^\circ$) for the eastern (J1418)
and the western (R2) source, respectively.
We produce and inspect the light curves of the sources, and find
some flares. We suspect that the standard flare-filtering ($\le$0.4\,$\rm s^{-1}$)
leaves a large number of particle-flare events unfiltered (Fig.~\ref{fig:fig2} left).
This may be a concern for timing studies because
backgrounds cannot be directly subtracted in timing analyses (e.g., Fourier analysis
and epoch-folding techniques). We therefore
inspect the flare light curve (Fig.~\ref{fig:fig2} left),
and find that we can better remove the flare backgrounds
by adjusting the flare cut to a lower value.

	After visual inspection of the flare light curve,
we change the flare cut to $\le$0.15\,$\rm s^{-1}$ (see below for different cuts).
This reduces the number of events in the source region of J1418 by $\approx$30\%.
We then fold the event arrival times
using the {\it Fermi}-LAT timing solution \citep[][]{krjs+15} and find that
pulsation of J1418 is detected significantly ($p=2\times 10^{-7}$ for $H=39$)
but not for R2;
the folded X-ray pulse profile of J1418 shows a sharp peak at $\phi\approx 0.1$
and another smaller (and less clear) one at $\phi\approx 0.6$, similar to 
that in Figure~\ref{fig:fig2} right
and a previously-reported one.\footnote{https://cxc.harvard.edu/cdo/snr09/pres/Roberts\_Mallory.pdf}
These peak phases are similar to those of the GeV
peaks.\footnote{https://www.slac.stanford.edu/$\sim$kerrm/fermi\_pulsar\_timing/}

\newcommand{\markaa}{\tablenotemark{a}}
\begin{table}
\vspace{-0.0in}
\begin{center}
\caption{Detection significance of the pulsation for various event selections}
\label{ta:ta2}
\vspace{-0.02in}
\scriptsize{
\begin{tabular}{lcccc} \hline\hline
Flare cut & Energy band & Region shift        & $H$-value & Cts/Exposure\\
          & (keV)       & ($''$/$''$)\markaa  &           & (/ks)       \\ \hline
0.03      & 1--10       & 0/0                 & 32.07     & 741/48   \\
0.07      & 1--10       & 0/0                 & 40.50     & 1064/64  \\ 
0.1       & 1--10       & 0/0                 & 46.26     & 1130/67  \\
0.2       & 1--10       & 0/0                 & 43.92     & 1430/74  \\
0.3       & 1--10       & 0/0                 & 40.05     & 1720/79  \\
0.4       & 1--10       & 0/0                 & 36.41     & 1830/81  \\ \hline
0.15      & 0.5--10     & 0/0                 & 36.06     & 1370/71  \\
0.15      & 1--10       & 0/0                 & 38.69     & 1300/71  \\
0.15      & 1.5--10     & 0/0                 & 33.68     & 1200/71  \\ \hline
0.15      & 1--8        & 0/0                 & 38.48     & 1210/71  \\
0.15      & 1--6        & 0/0                 & 32.25     & 1036/71  \\ \hline
0.15      & 1--10       & 3/0                 & 41.25     & 1290/71  \\
0.15      & 1--10       & $-$3/0              & 28.22     & 1320/71  \\
0.15      & 1--10       & 0/3                 & 25.28     & 1290/71  \\
0.15      & 1--10       & 0/$-$3              & 43.65     & 1280/71  \\ \hline
\end{tabular}}
\end{center}
\footnotesize{
$^{\rm a}${R.A./Decl.}}
\vspace{-0.5 mm}
\end{table}

We further check to see if the results change significantly when using different
event selections: energy band, source region, or flare cut (Table~\ref{ta:ta2}).
The $H$ value for the pulsation varies between 25 and 46
($p=4\times 10^{-5}$--$9\times 10^{-9}$) when
varying the energy band (e.g., lower bound 0.5--1.5\,keV and upper bound 6--10\,keV),
source-region center (by 3$''$ in all directions), or the flare cut value (0.03--0.4);
the detection becomes less significant (lower $H$ values) for source regions that are close to R2,
narrower energy bands or non-optimal flare-cut values.
Nevertheless, all of these selections result in a significant detection, and hence the detection
of the pulsation in J1418 is robust.
However, we note that using a large region (e.g., $R>20''$) and/or a narrower energy range 
(e.g., 3--9\,keV)
reduces the signal-to-noise ratio and makes the detection insignificant.

Because the LAT timing solution was constructed by using large time bins compared
to the exposure of the X-ray data, the solution may not properly account for timing noise
effects at the epoch of the X-ray observation and hence may not be optimal. We therefore
try to find a better frequency by carrying out $H$ tests \citep[][]{drs89} near the LAT solution
$f=9.0438674 - 9.0438762\rm \ s^{-1}$ and
find a sharp peak in the $H$ plot (Fig.~\ref{fig:fig2} middle)
at $f=9.0438699\rm \ s^{-1}$ with $H=40$ (reference epoch MJD~54883), corresponding to $p=1\times 10^{-7}$.
This result is robust to a small change of event selection (energy band, region, or flare cut)
Note that we cannot determine frequency derivatives because of the low count statistics.
The resulting pulse profile is displayed in Figure~\ref{fig:fig2} right.

\subsection{The pulsar spectral analyses}
\label{sec:sec2_3}
	J1418 was suggested to be a soft-gamma pulsar candidate \citep[][]{kh15},
and then its X-ray spectrum is presumably hard. Now that we clearly identified
it as a pulsar, we are able to measure its pulsed spectrum as well as phase-summed one.

\newcommand{\marka}{\tablenotemark{a}}
\begin{table*}[t]
\vspace{-0.0in}
\begin{center}
\caption{Power-law fit results for the pulsar and the extended emission}
\label{ta:ta1}
\vspace{-0.05in}
\scriptsize{
\begin{tabular}{lccccccc} \hline\hline
Data     & Instrument  & Energy range   & $N_{\rm H}$      & $\Gamma$      &  $F_{\rm X}$ & $lstat$/dof\\
         & & (keV) & ($10^{22}\rm \ cm^{-2}$) &  & ($10^{-13}\rm \ erg\ s^{-1}\ cm^{-2}$) &   & \\ \hline
PSR phase summed  & {\it Chandra} & 0.5--10 & $2.8\pm0.7$ & $1.5\pm0.4$   & $1.5\pm0.3$   & 31/41 \\
PSR pulsed        & {\it XMM}/PN  & 0.5--10 & $2.8$\marka & $1.0\pm0.6$   & $1.3\pm0.3$   & 54/66 \\
Extended emission & {\it Chandra} & 0.5--10 & $2.8$\marka & $1.8\pm0.3$   & $0.7\pm0.1$   & 23/16 \\ \hline
\end{tabular}}
\end{center}
\vspace{-0.5 mm}
\footnotesize{
$^{\rm a}${Fixed at the value measured for the phase-summed spectrum.}}
\end{table*}

	For a phase-summed spectral analysis,
{\it XMM-Newton} data (both MOS and PN) are not very useful because
the pulsar is relatively faint and so contamination from emission of
the bright surrounding region (e.g., R2 in Fig.~\ref{fig:fig1} right) is a concern.
We therefore use only the high spatial-resolution {\it Chandra} data for
the phase-summed spectral analysis.
We extract source events within an $R=2''$ circle
and background events using an $R=2.1-5''$ annular region; the latter is selected
so as to properly take into account backgrounds from the possible
extended emission (Fig.~\ref{fig:fig1} and Section~\ref{sec:sec2_4}).
Corresponding response files
are calculated with the {\tt spec\_extract} tool of CIAO.
We group the spectrum to have at least 5 events in each spectral bin,
fit the 0.5--10\,keV spectrum with an absorbed power-law model
in XSPEC V12.10.1 employing {\it lstat} \citep[][]{l92} because of paucity of counts.
We find that the power-law model with
$N_{\rm H}=2.8\pm0.7\times 10^{22}\rm \ cm^{-2}$,
$\Gamma=1.5\pm0.4$, and
0.5--10\,keV flux $F_{\rm X}=1.5\pm 0.3\times 10^{-13}\rm \ erg\ cm^{-2}\ s^{-1}$
adequately describes the data ($lstat$/dof=31/41).
The results are summarized in Table~\ref{ta:ta1}.
A blackbody model also explains the data but the best-fit temperature appears
to be too high ($kT=1.5$\,keV). So we do not report the results.
Note that these results obtained with the 70-ks {\it Chandra} data
are qualitatively similar to those reported
by \citet{nrr05} based on 10-ks {\it Chandra} and 25-ks {\it XMM-Newton} data,
but quantitative comparison is not possible because detailed information
on the analysis is missing in that work.

\begin{figure}
\centering
\hspace{0.0 mm}
\includegraphics[width=3.25 in]{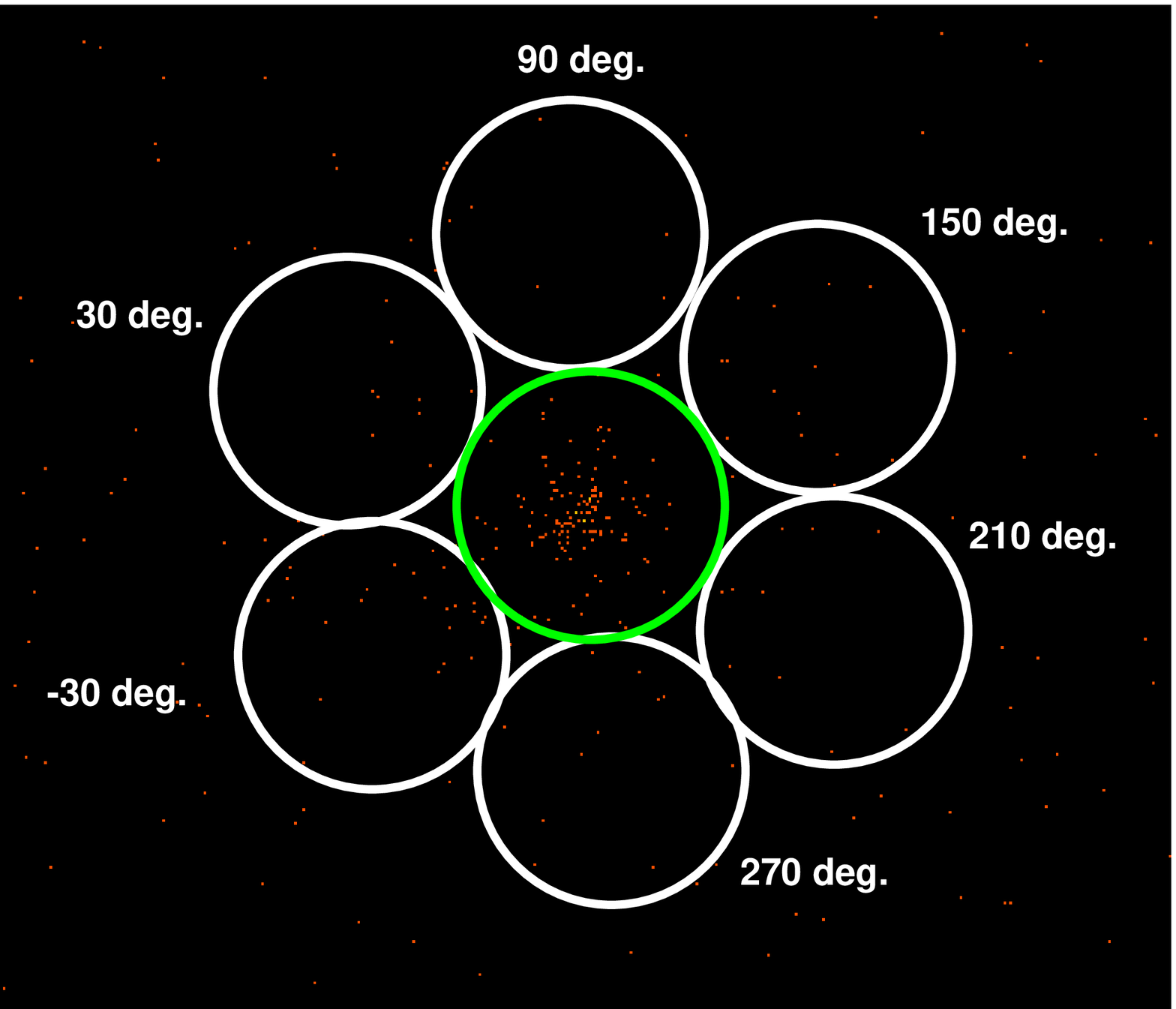}
\includegraphics[width=3.25 in]{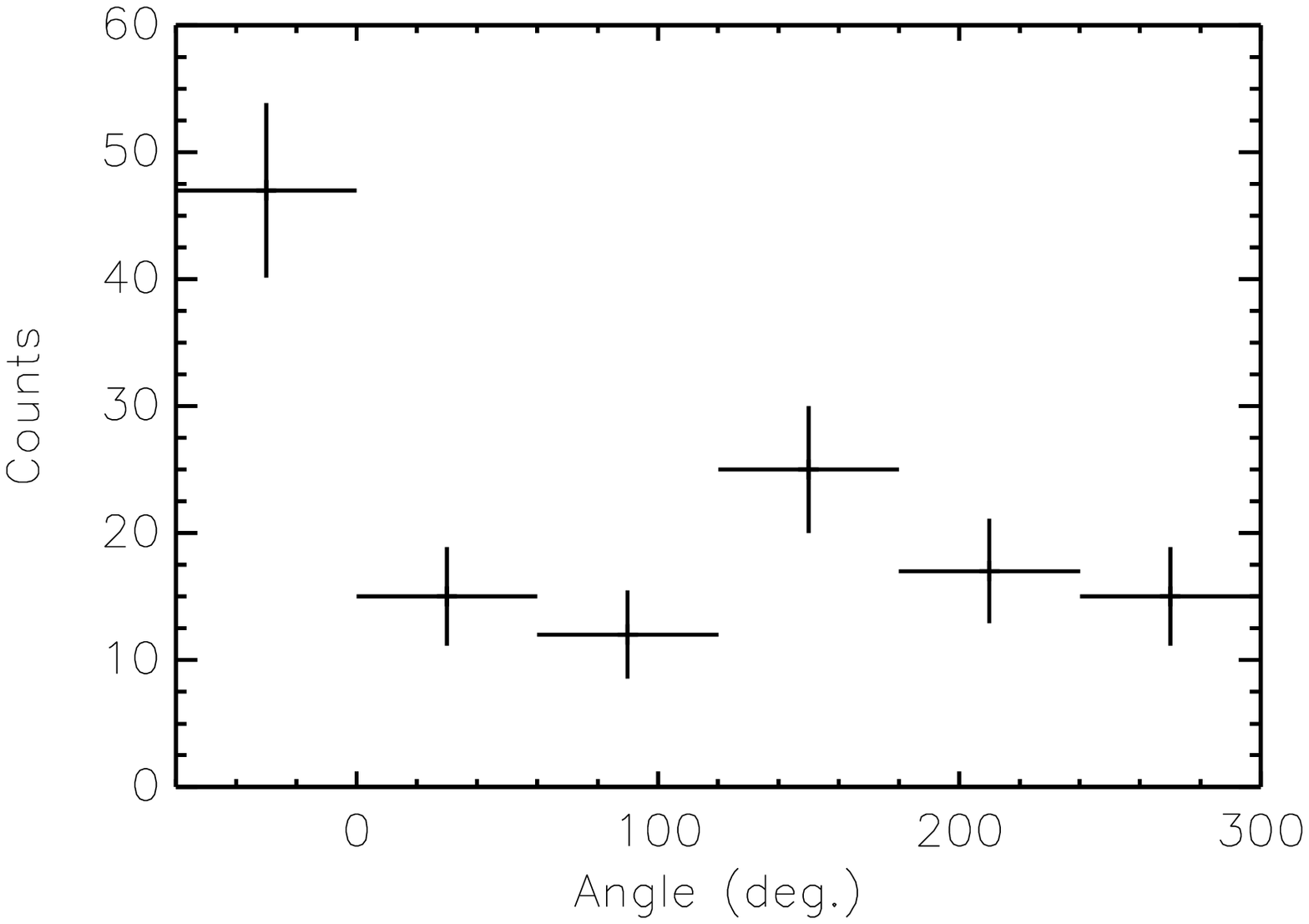}
\figcaption{
{\it Top}: A 1--10\,keV {\it Chandra} image of J1418 vicinity with regions surrounding the pulsar (green circle).
{\it Bottom}: Numbers of events contained within $R=2''$ circles (white circles in the top panel).
The angle is defined from east to north.
\label{fig:fig3}
}
\vspace{0mm}
\end{figure}

	We next measure the pulsed-spectrum using the {\it XMM-newton}/PN data.
Although there is a {\it Fermi}-LAT timing solution, we use our own
solution because the LAT solution was constructed by using
relatively large time bins and may not adequately account for
timing noise with the timescale of the 120-ks {\it XMM-newton} exposure.
The pulse profile seems to consist of two peaks around $\phi=0.1$ and $0.6$, but the 
second peak is less significant (Fig.~\ref{fig:fig2} right).
We therefore focus on the first peak here.
We select phase intervals of $\phi=0-0.2$ and $\phi=0.2-1.0$
for the pulsed spectrum and the unpulsed background, respectively.
We extract the spectra using an $R=10''$ circular region in the {\it XMM-Newton}/PN data,
and compute the response files using the {\tt rmfgen} and the {\tt arfgen} tasks of SAS.
We group the pulsed spectrum to have at least 5 events per bin
and fit the 0.5--10\,keV spectrum with an 
absorbed power law holding $N_{\rm H}$ fixed at
the {\it Chandra}-measured value for the phase-summed spectrum.
The model describes the spectrum well ($lstat$/dof=54/66) with 
$\Gamma=1.0\pm0.6$ and the 0.5--10\,keV flux
$F_{\rm X}=1.3\pm0.3\times 10^{-13}\rm \ erg\ cm^{-2}\ s^{-1}$.

\subsection{Image analysis}
\label{sec:sec2_4}
	We inspect multi-epoch images obtained with the {\it Chandra} and {\it XMM-Newton}
observations.
Figure~\ref{fig:fig1} shows 0.5--10\,keV {\it Chandra} and {\it XMM-Newton} images
of the field. In these images, we find that J1418 is persistently detected.
However, the source R3 in the {\it Chandra} image is not present in the {\it XMM-Newton} data.
Instead, another source R2 is detected at $\sim$10$''$ north to R3,
meaning that these two sources R2 and R3 are variable.
Note that considering agreement between {\it Chandra} and {\it XMM-Newton} positions of other
sources in the field, this is not due to the positional inaccuracy of the observatories.
Since R2 and R3 are highly variable, they are not likely to be a pulsar.

	Focusing on J1418, we find extended emission around it
in the south-east to north-west direction
($\approx$5$''$; Fig.~\ref{fig:fig1} left and Fig.~\ref{fig:fig3}) with the eastern part
being slightly brighter. In order to verify this, we count the number
of 1--10\,keV events within six $R=2''$ circles surrounding
the J1418 region (Fig.~\ref{fig:fig3} top),
and show the angular distribution of the counts in Figure~\ref{fig:fig3} bottom.
The angle ($\theta$) is defined from east to north,
so that the $\theta\approx-30^\circ$ zone corresponds to
the south-east region. This region
contains $47\pm7$ events as compared to $17\pm2$, the average of the other
regions; the difference is $4.7\sigma$ significant. The north-west region ($\theta=150^\circ$)
also seems to have more counts than the other four lower-count zones ($2\sigma$ significant).
A formal $\chi^2$ test rules out a flat distribution with $p=2\times 10^{-5}$.

	We confirm the source extendedness
using the {\tt srcextent} tool of CIAO.
We generate point source events (i.e., point spread function; PSF) using
a {\it MARX} simulation\footnote{https://space.mit.edu/cxc/marx/}
appropriate to the observation parameters and the source spectrum we measured (Table~\ref{ta:ta1}).
We then run the {\tt srcextent} script with the observed event and the simulated PSF files
on various spatial scales $R=2-10''$, and find that the source is extended at 90\% confidence
for $R\ge3''$ scales. At $R<3''$, the event distribution is not significantly
different from the PSF at the 90\% confidence level.
We also check to see if the extended emission is produced by a faint source adjacent
to J1418 using the source detection tool {\tt wavdetect} of CIAO. It properly detects nearby
point sources (e.g., R3 in Fig.~\ref{fig:fig1}), but none in the region surrounding J1418.
Hence the extended emission seems not to arise from a faint point source.

	In order to measure the spectrum of the extended emission, we extract
source events within a $3''\times 5''$ ellipse excluding the central $R=2''$ region,
and background events within a $R=50''$ circle $2.5'$ north to the pulsar.
Response files for the extended emission are generated with the {\tt spec\_extract}
tool of CIAO. We group the source spectrum to have at least 5 events per spectral bin
and fit the spectrum with an absorbed power-law model holding $N_{\rm H}$ fixed at
$2.8 \times 10^{22}\rm \ cm^{-2}$. The power-law model describes
the spectrum well ($lstat$/dof=23/16) having $\Gamma=1.8\pm0.3$ and 0.5--10\,keV flux
$F_{\rm X}=7\pm1\times10^{-14} \rm \ erg\ cm^{-2}\ s^{-1}$ (Table~\ref{ta:ta1}).

\section{Discussion and Conclusions}
\label{sec:sec4}
	We found significant X-ray pulsation of J1418 in the
Rabbit PWN, thereby confirming that J1418 is the X-ray
counterpart of the gamma-ray pulsar PSR~J1418$-$6058.
Our image analysis revealed extended emission around the pulsar
in the south-east to north-west direction with a position angle $30^\circ$ from
west to north. The extended emission may be
a torus or jets often seen around pulsars in young PWNe.
We note that similar and qualitative results were reported
previously.\footnote{https://cxc.harvard.edu/cdo/snr09/pres/Roberts\_Mallory.pdf}
We also find that the other X-ray sources in the Rabbit PWN are
variable and hence are not pulsars. These findings strongly suggest
that J1418 is the power source of the Rabbit PWN.
The pulsar's X-ray spectra appear to be hard with $\Gamma\approx1$ similar to
other soft-gamma pulsars.

	The {\it XMM-Newton}/PN small-window data provide sufficient
timing resolution to search for the 110\,ms pulsation of J1418.
Searches for its X-ray pulsation were attempted in the past
with the same {\it XMM-Newton}/PN data, but the pulsation was not significantly
detected \citep[e.g.,][]{kh15}.
This is probably because they used a large region ($R=15''$), narrow energy bands
(0.3--2\,keV or 2--10\,keV), and/or the standard flare cut ($\le$0.4\,$\rm s^{-1}$).
Here we showed that optimizing the event selection in {\it XMM-Newton}
data analyses can increase sensitivity for pulsation search and result in a significant detection.

	In the high-resolution {\it Chandra} image, we found
extended emission around J1418. Similar features are
seen in young X-ray bright PWNe and interpreted as a torus corresponding to
termination shock or bipolar jets
\citep[e.g.,Crab, MSH~15$-$5{\sl 2}, and 3C~58;][]{mrha+15,amrk+14,shvm04}.
Although it is not clear whether the extended emission around J1418 is
jets or a torus, the emission is brighter in the eastern side,
which could be due to the Doppler boosting \citep{nr04} of collimated bipolar outflow,
suggesting that the feature perhaps corresponds to jets. If so,
we can estimate the jet viewing angle to be $\zeta=67^\circ$ for the jet-to-counterjet
count ratio of $\sim$3 (Fig.~\ref{fig:fig3} bottom)
and the photon index $\Gamma=1.8$ (Table~\ref{ta:ta1}), using a Doppler boost formula
\citep[e.g.,][]{nr04} for an assumed bulk flow speed of $\beta_{\rm jet}=0.5$.
Further confirmation can be made by comparing the pulsar emission geometry
inferred from gamma-ray profile fitting \citep[][]{hma98,rw10} with
that inferred from torus fitting \citep[e.g.,][]{nr04}.

The timing properties of J1418, spin-down luminosity
$L_{\rm sd}=5\times10^{36}\rm \ erg\ s^{-1}$,
the characteristic age $\tau_c=10$\,kyr, and the magnetic-field strength
$B_s=4\times 10^{12}$\,G \citep[][]{krjs+15},
are within the ranges for those of soft-gamma pulsars. Hence,
the source was suggested to be a soft-gamma pulsar candidate previously \citep[][]{kh15}.
We found that the X-ray spectra of the pulsar are hard similar to those of
soft-gamma pulsars ($\Gamma_X\approx 1$), confirming
the previous suggestion. Hence J1418 may add to the list of soft-gamma pulsars.

	With the current data, it is hard to measure properties of J1418
well. Accurate measurements of the X-ray pulse profile and spectra
can help to understand the pulsar and the Rabbit PWN.
In particular, identifying the extended emission as a torus or jets
can be useful for understanding the energy injection to the Rabbit PWN.
Since the pulsar has hard X-ray spectra, {\it NuSTAR} \citep[][]{hcc+13} observations
can easily detect the pulsations \citep[e.g., as in PSR~J0205$+$6449;][]{a19}
and characterize the pulse profile and emission spectrum. So further
studies with {\it NuSTAR} are warranted.

\bigskip
\bigskip

\acknowledgments

We thank Matthew Kerr for discussions on the timing works.
This research was supported by Basic Science Research Program through
the National Research Foundation of Korea (NRF)
funded by the Ministry of Science, ICT \& Future Planning (NRF-2017R1C1B2004566).

\vspace{5mm}
\facilities{CXO, XMM}
\software{HEAsoft (v6.24; HEASARC 2014), CIAO \citep[v4.11; ][]{fmab+06},
XMM-SAS \citep[v20180620; ][]{xmmsas17}, XSPEC \citep[][]{a96}}


\bibliographystyle{apj}
\bibliography{MAGNETAR,GBINARY,BLLacs,PSRBINARY,PWN,STATISTICS,FERMIBASE,COMPUTING,INSTRUMENT,ABSORB,PULSARS}

\begin{thebibliography}{}
\expandafter\ifx\csname natexlab\endcsname\relax\def\natexlab#1{#1}\fi

\bibitem[{{Abdo} {et~al.}(2013){Abdo}, {Ajello}, {Allafort}, {Baldini},
  {Ballet}, {Barbiellini}, {Baring}, {Bastieri}, {Belfiore}, {Bellazzini}, \&
  et~al.}]{fermi2PC}
{Abdo}, A.~A., {Ajello}, M., {Allafort}, A., {et~al.} 2013, \apjs, 208, 17

\bibitem[{{Aharonian} {et~al.}(2006){Aharonian}, {Akhperjanian}, {Bazer-Bachi},
  {Beilicke}, {Benbow}, {Berge}, {Bernl{\"o}hr}, {Boisson}, {Bolz}, {Borrel},
  {Braun}, {Brown}, {B{\"u}hler}, {B{\"u}sching}, {Carrigan}, {Chadwick},
  {Chounet}, {Cornils}, {Costamante}, {Degrange}, {Dickinson},
  {Djannati-Ata{\"\i}}, {O'C. Drury}, {Dubus}, {Egberts}, {Emmanoulopoulos},
  {Espigat}, {Feinstein}, {Ferrero}, {Fiasson}, {Fontaine}, {Funk}, {Funk},
  {F{\"u}{\ss}ling}, {Gallant}, {Giebels}, {Glicenstein}, {Goret},
  {Hadjichristidis}, {Hauser}, {Hauser}, {Heinzelmann}, {Henri}, {Hermann},
  {Hinton}, {Hoffmann}, {Hofmann}, {Holleran}, {Horns}, {Jacholkowska}, {de
  Jager}, {Kendziorra}, {Kh{\'e}lifi}, {Komin}, {Konopelko}, {Kosack},
  {Latham}, {Le Gallou}, {Lemi{\`e}re}, {Lemoine-Goumard}, {Lohse}, {Martin},
  {Martineau-Huynh}, {Marcowith}, {Masterson}, {Maurin}, {McComb}, {de
  Naurois}, {Nedbal}, {Nolan}, {Noutsos}, {Orford}, {Osborne}, {Ouchrif},
  {Panter}, {Pelletier}, {Pita}, {P{\"u}hlhofer}, {Punch}, {Raubenheimer},
  {Raue}, {Rayner}, {Reimer}, {Reimer}, {Ripken}, {Rob}, {Rolland }, {Rowell},
  {Sahakian}, {Santangelo}, {Saug{\'e}}, {Schlenker}, {Schlickeiser},
  {Schr{\"o}der}, {Schwanke}, {Schwarzburg}, {Shalchi}, {Sol}, {Spangler},
  {Spanier}, {Steenkamp}, {Stegmann}, {Superina}, {Tavernet}, {Terrier},
  {Th{\'e}oret}, {Tluczykont}, {van Eldik}, {Vasileiadis}, {Venter}, {Vincent},
  {V{\"o}lk}, {Wagner}, \& {Ward}}]{hessrabbit2006}
{Aharonian}, F., {Akhperjanian}, A.~G., {Bazer-Bachi}, A.~R., {et~al.} 2006,
  \aap, 456, 245

\bibitem[{{An}(2019)}]{a19}
{An}, H. 2019, \apj, 876, 150

\bibitem[{{An} {et~al.}(2014){An}, {Madsen}, {Reynolds}, {Kaspi}, {Harrison},
  {Boggs}, {Christensen}, {Craig}, {Fryer}, {Grefenstette}, {Hailey}, {Mori},
  {Stern}, \& {Zhang}}]{amrk+14}
{An}, H., {Madsen}, K.~K., {Reynolds}, S.~P., {et~al.} 2014, \apj, 793, 90

\bibitem[{{Arnaud}(1996)}]{a96}
{Arnaud}, K.~A. 1996, in Astronomical Society of the Pacific Conference Series,
  Vol. 101, Astronomical Data Analysis Software and Systems V, ed. G.~H.
  {Jacoby} \& J.~{Barnes}, 17

\bibitem[{{Atwood} {et~al.}(2009){Atwood}, {Abdo}, {Ackermann}, {Althouse},
  {Anderson}, {Axelsson}, {Baldini}, {Ballet}, {Band}, {Barbiellini}, \&
  et~al.}]{fermimission}
{Atwood}, W.~B., {Abdo}, A.~A., {Ackermann}, M., {et~al.} 2009, \apj, 697, 1071

\bibitem[{{Coti Zelati} {et~al.}(2019){Coti Zelati}, {Torres}, {Li}, \&
  {Vigano}}]{ctlv19}
{Coti Zelati}, F., {Torres}, D.~F., {Li}, J., \& {Vigano}, D. 2019, arXiv
  e-prints, arXiv:1912.03953

\bibitem[{{de Jager} {et~al.}(1989){de Jager}, {Raubenheimer}, \&
  {Swanepoel}}]{drs89}
{de Jager}, O.~C., {Raubenheimer}, B.~C., \& {Swanepoel}, J.~W.~H. 1989, \aap,
  221, 180

\bibitem[{{Fruscione} {et~al.}(2006){Fruscione}, {McDowell}, {Allen},
  {Brickhouse}, {Burke}, {Davis}, {Durham}, {Elvis}, {Galle}, {Harris},
  {Huenemoerder}, {Houck}, {Ishibashi}, {Karovska}, {Nicastro}, {Noble},
  {Nowak}, {Primini}, {Siemiginowska}, {Smith}, \& {Wise}}]{fmab+06}
{Fruscione}, A., {McDowell}, J.~C., {Allen}, G.~E., {et~al.} 2006, in
  \procspie, Vol. 6270, Society of Photo-Optical Instrumentation Engineers
  (SPIE) Conference Series, 62701V

\bibitem[{{Gabriel}(2017)}]{xmmsas17}
{Gabriel}, C. 2017, in The X-ray Universe 2017, 84

\bibitem[{{Harding}(2013)}]{harding13}
{Harding}, A.~K. 2013, Frontiers of Physics, 8, 679

\bibitem[{{Harding} \& {Muslimov}(1998)}]{hma98}
{Harding}, A.~K., \& {Muslimov}, A.~G. 1998, \apj, 500, 862

\bibitem[{{Harrison} {et~al.}(2013){Harrison}, {Craig}, {Christensen},
  {Hailey}, {Zhang}, {Boggs}, {Stern}, {Cook}, {Forster}, {Giommi},
  {Grefenstette}, {Kim}, {Kitaguchi}, {Koglin}, {Madsen}, {Mao}, {Miyasaka},
  {Mori}, {Perri}, {Pivovaroff}, {Puccetti}, {Rana}, {Westergaard}, {Willis},
  {Zoglauer}, {An}, {Bachetti}, {Barri{\`e}re}, {Bellm}, {Bhalerao},
  {Brejnholt}, {Fuerst}, {Liebe}, {Markwardt}, {Nynka}, {Vogel}, {Walton},
  {Wik}, {Alexander}, {Cominsky}, {Hornschemeier}, {Hornstrup}, {Kaspi},
  {Madejski}, {Matt}, {Molendi}, {Smith}, {Tomsick}, {Ajello}, {Ballantyne},
  {Balokovi{\'c}}, {Barret}, {Bauer}, {Blandford}, {Brandt}, {Brenneman},
  {Chiang}, {Chakrabarty}, {Chenevez}, {Comastri}, {Dufour}, {Elvis}, {Fabian},
  {Farrah}, {Fryer}, {Gotthelf}, {Grindlay}, {Helfand}, {Krivonos}, {Meier},
  {Miller}, {Natalucci}, {Ogle}, {Ofek}, {Ptak}, {Reynolds}, {Rigby},
  {Tagliaferri}, {Thorsett}, {Treister}, \& {Urry}}]{hcc+13}
{Harrison}, F.~A., {Craig}, W.~W., {Christensen}, F.~E., {et~al.} 2013, \apj,
  770, 103

\bibitem[{{Kaspi} \& {Beloborodov}(2017)}]{kb17}
{Kaspi}, V.~M., \& {Beloborodov}, A.~M. 2017, \araa, 55, 261

\bibitem[{{Kennel} \& {Coroniti}(1984)}]{kc84a}
{Kennel}, C.~F., \& {Coroniti}, F.~V. 1984, \apj, 283, 694

\bibitem[{{Kerr} {et~al.}(2015){Kerr}, {Ray}, {Johnston}, {Shannon}, \&
  {Camilo}}]{krjs+15}
{Kerr}, M., {Ray}, P.~S., {Johnston}, S., {Shannon}, R.~M., \& {Camilo}, F.
  2015, \apj, 814, 128

\bibitem[{{Kuiper} \& {Hermsen}(2015)}]{kh15}
{Kuiper}, L., \& {Hermsen}, W. 2015, \mnras, 449, 3827

\bibitem[{{Lamb} \& {Macomb}(1997)}]{lm97}
{Lamb}, R.~C., \& {Macomb}, D.~J. 1997, \apj, 488, 872

\bibitem[{{Loredo}(1992)}]{l92}
{Loredo}, T.~J. 1992, in Statistical Challenges in Modern Astronomy, ed. E.~D.
  {Feigelson} \& G.~J. {Babu}, 275--297

\bibitem[{{Madsen} {et~al.}(2015){Madsen}, {Reynolds}, {Harrison}, {An},
  {Boggs}, {Christensen}, {Craig}, {Fryer}, {Grefenstette}, {Hailey},
  {Markwardt}, {Nynka}, {Stern}, {Zoglauer}, \& {Zhang}}]{mrha+15}
{Madsen}, K.~K., {Reynolds}, S., {Harrison}, F., {et~al.} 2015, \apj, 801, 66

\bibitem[{{Ng} {et~al.}(2005){Ng}, {Roberts}, \& {Romani}}]{nrr05}
{Ng}, C.~Y., {Roberts}, M. S.~E., \& {Romani}, R.~W. 2005, \apj, 627, 904

\bibitem[{{Ng} \& {Romani}(2004)}]{nr04}
{Ng}, C.~Y., \& {Romani}, R.~W. 2004, \apj, 601, 479

\bibitem[{{Roberts} {et~al.}(1999){Roberts}, {Romani}, {Johnston}, \&
  {Green}}]{rrjg99}
{Roberts}, M. S.~E., {Romani}, R.~W., {Johnston}, S., \& {Green}, A.~J. 1999,
  \apj, 515, 712

\bibitem[{{Romani} \& {Watters}(2010)}]{rw10}
{Romani}, R.~W., \& {Watters}, K.~P. 2010, \apj, 714, 810

\bibitem[{{Sironi} {et~al.}(2015){Sironi}, {Keshet}, \& {Lemoine}}]{skl15}
{Sironi}, L., {Keshet}, U., \& {Lemoine}, M. 2015, \ssr, 191, 519

\bibitem[{{Slane} {et~al.}(2004){Slane}, {Helfand}, {van der Swaluw}, \&
  {Murray}}]{shvm04}
{Slane}, P., {Helfand}, D.~J., {van der Swaluw}, E., \& {Murray}, S.~S. 2004,
  \apj, 616, 403

\end{thebibliography}

\end{document}